\documentclass[aps,showpacs,amsmath,amssymb,12pt]{revtex4}
\usepackage{latexsym}
\usepackage{bm}

 \def\G{\Gamma}

\def\d{\delta} 
  
\def\m{\mu} 
\def\n{\nu}

\def\p{\phi} \def\P{\Phi}
\def\x{\xi}
\def\z{\zeta}
\def\vp{\varphi}
 
\def\t{\tau} 
  \def\S{\Sigma}

\def\io{\iota} 
      
\def\we{\wedge}

\def\be{\begin{equation}}
\def\ee{\end{equation}}

\def\HollowBox #1#2{{\dimen0=#1 \advance\dimen0 by -#2
       \dimen1=#1 \advance\dimen1 by #2
        \vrule height #1 depth #2 width #2
        \vrule height 0pt depth #2 width #1
        \llap{\vrule height #1 depth -\dimen0 width \dimen1} 
       \hskip -#2
       \vrule height #1 depth #2 width #2}}
\def\BOX{\HollowBox{.100in}{.010in}}

\begin{document}

\title{Gravitational charges of \emph{transverse} asymptotically 
$AdS$ spacetimes} 

\author{Hakan Cebeci}
\email{hcebeci@anadolu.edu.tr}
\affiliation{Anadolu University, Department of Physics, Yunus Emre
Campus, 26470, Eski{\c s}ehir, Turkey}

\author{{\"O}zg{\"u}r Sar{\i}o\u{g}lu}
\email{sarioglu@metu.edu.tr}
\affiliation{Department of Physics, Faculty of Arts and Sciences,\\
             Middle East Technical University, 06531, Ankara, Turkey}

\author{Bayram Tekin}
\email{btekin@metu.edu.tr}
\affiliation{Department of Physics, Faculty of Arts and Sciences,\\
             Middle East Technical University, 06531, Ankara, Turkey}

\date{\today}

\begin{abstract} 
Using Killing-Yano symmetries, we construct conserved charges of
spacetimes that asymptotically approach to the flat or Anti-de Sitter spaces 
\emph{only in certain} directions. In $D$ dimensions, this allows one to 
define gravitational charges (such as mass and angular momenta densities)
of $p$-dimensional branes/solitons or any other extended objects
that curve the transverse space into an asymptotically flat or $AdS$ one.
Our construction answers the question of what kind of charges the
antisymmetric Killing-Yano tensors lead to.
\end{abstract}


\maketitle

%\newpage

\section{\label{intro} Introduction}

The use of background symmetries (such as ``Killing vector symmetries") 
for defining conserved charges in gravitational theories - those arising 
from the extremization of an action or those endowed with ``Bianchi 
identities" - has borne much fruit. Two immediate consequences of such 
definitions are: the background gauge invariance of the charges and the 
vanishing of the background charge. One then computes the charges of 
those spacetimes that asymptotically approach the background. It is 
important to note that these charges are expressible as surface integrals 
at the spatial boundary of spacetime. What is implicit in this construction 
is that the asymptotic structure of the spacetime is either flat or 
$AdS$, be it locally or not, and the integrations are carried out on 
$(D-2)$-dimensional ``spheres'' or tori (or the products of these) (see 
e.g. \cite{adm, abdes, des1}). 

These definitions allow for the calculation of the charges of spacetimes
(such as black holes, localized solitons and so on) that approach to flat 
or $AdS$ geometries in ``all dimensions". However, a naive application of 
these charge definitions to spacetimes that approach to flat or $AdS$ 
spaces only in certain \emph{transverse} directions immediately leads to 
trivial divergences. For example, the ``mass'' of an extended solution,
such as a $p$-brane, which is infinite in specific directions and curves 
the \emph{transverse} space into flat or $AdS$ geometry, would be divergent.
Thus, the question whether one can construct a reasonable charge 
definition which only considers the transverse part of such spacetimes is 
of relevance.

An important step in this direction, for transverse asymptotically flat 
case, was taken by Kastor and Traschen \cite{kt1} who used background 
antisymmetric Killing-Yano tensors, instead of background Killing vectors. 
Quite interestingly, unlike the usual Killing charges (such as energy and 
angular momentum) which are built out of the linearization of the Einstein's 
equations, Killing-Yano charges are constructed from a new antisymmetric 
current which makes no reference to Einstein's equations \cite{kt1}. Kastor 
and Traschen termed these ``new" conserved quantities as Y-ADM charges. 
For rank $n$ Killing-Yano tensors, they worked out the case when the 
transverse part of the spacetime is asymptotically flat. In a follow-up paper
\cite{kt2}, they discussed the conditions necessary for the positivity of 
Y-ADM ``mass". One very important result of their work is that even though 
they have constructed these Y-ADM charges from new antisymmetric currents 
and Killing-Yano tensors, making no reference to Einstein's equations, they 
obtained the usual ADM expression, except that the integrations are now to be 
carried out only in transverse directions instead of the usual 
$(D-2)$-dimensional spheres or tori. Thus, for example, as far as the 
``mass-like" Y-ADM charge is concerned, what they ended up with is just 
the mass density of, say, a $p$-brane that either extends to infinity in 
$p$ spatial directions or is periodic in certain directions. 

We note that Killing-Yano tensors, unlike the Killing vectors, are not 
related to the isometries of the metric; thus they do not lead to the
already known symmetries and conserved quantities. However, in different
contexts they have been shown to yield various nontrivial constants of
motion which are related to the separability of the Hamilton-Jacobi and
wave equations. The most important application of this is the Kerr-Newman
metric in four dimensions \cite{carter}. [See Gibbons et al. \cite{gibb}
for more details and the connection between Killing-Yano tensors and a 
nontrivial supersymmetry which is related to the square root of a constant
of motion (not the usual Hamiltonian) derived from a Killing-Yano tensor.]
Here we shall give another use of these tensors. The main point is that, 
if a spacetime admits such objects, one can construct new conserved 
currents \cite{kt1} whose physical meaning we try to interpret here.

One important question which was mentioned by Kastor and Traschen is 
finding the surface integral expressions of the Killing-Yano charges 
for transverse asymptotic $AdS$ spacetimes. Our main task in this paper 
will be to carry out in full detail the construction of these conserved 
quantities for such spacetimes admitting rank $n$ Killing-Yano tensors. 
We shall call these Y-AD (Yano-Abbott-Deser) charge densities. In the
next section, which is the bulk of our paper, we give the construction
of these and present an example of the long axisymmetric Weyl rod.

\section{\label{kyc} Killing-Yano charges of transverse 
asymptotically $AdS$ spacetimes}

Similar to the construction of the background Killing charges for
asymptotically $AdS$ spacetimes \cite{abdes, des1}, here we will be 
interested in transverse asymptotically $AdS$ spacetimes that
asymptotically admit completely antisymmetric tensor fields which
play an analogous role to the usual background Killing symmetries.
For clarity of the discussion, we will first concentrate on the rank 
2 case in what follows. Later, we will generalize these results 
to the case of arbitrary rank $n$ Killing-Yano tensors.

Let us consider a $D$-dimensional spacetime $\bar{g}_{ab}$, which we
call `the background spacetime'. A completely antisymmetric rank $n$ 
tensor $\bar{f}_{a_1 a_2 \dots a_n}$ is a Killing-Yano tensor of this 
spacetime if it satisfies
\be
{\bar{\nabla}}_{a} \, \bar{f}_{b a_2 \dots a_n} 
+ {\bar{\nabla}}_{b} \, \bar{f}_{a a_2 \dots a_n} = 0 \, . \label{nyaneq}
\ee
Note that when rank equals 1, this reduces to the usual Killing vector 
definition. The spacetimes $g_{ab}$ whose Killing-Yano charges we will 
compute do not necessarily admit exact Killing-Yano tensors. However, 
we demand that the metric $g_{ab}$ can be asymptotically split into a 
background plus a perturbation: 
\be 
g_{ab} \equiv \bar{g}_{ab} + h_{ab} \, .
\ee 
We require that for certain transverse directions (whose number is
less than or equal to $(D-2)$), $g_{ab}$ asymptotically admits Killing-Yano 
tensors due to this splitting and the assumption that $h_{ab}$ vanishes 
sufficiently fast at the spatial infinity. Let us take the background
spacetime $\bar{g}_{ab}$ to be $AdS$, which obeys
\[ \bar{R}_{abcd} = \frac{2 \Lambda}{(D-1)(D-2)} \, 
(\bar{g}_{ac} \, \bar{g}_{bd} 
- \bar{g}_{ad} \, \bar{g}_{bc}) \, , \quad
\bar{R}_{ab} = \frac{2 \Lambda}{(D-2)} \, \bar{g}_{ab} \, , \quad
\bar{R} = \frac{2 \Lambda D}{(D-2)} \, . \]
In $D=4$, the $AdS$ spacetime admits 10 Killing-Yano tensors of rank 2
\cite{hoco, car}. For the number of Killing-Yano tensors of rank $n$ in
$D$ dimensions, see (2.4) of \cite{kt1}.

For the rank 2 case, one can easily prove the following identities,
which we freely make use of in the calculation of (\ref{yanc}) below:
\begin{eqnarray}
{\bar{\nabla}}_{a} \, \bar{f}^{ab} & = & 0 \, , \qquad
{\bar{\nabla}}_{a} \, \bar{f}_{bc} = {\bar{\nabla}}_{b} \, \bar{f}_{ca} 
= {\bar{\nabla}}_{c} \, \bar{f}_{ab} \, , \label{ide2} \\
{\bar{\nabla}}_{d} \, {\bar{\nabla}}_{a} \, \bar{f}_{bc} & = & 
\frac{2 \Lambda}{(D-1)(D-2)} (\bar{g}_{da} \, \bar{f}_{cb} + 
\bar{g}_{db} \, \bar{f}_{ac} + \bar{g}_{dc} \, \bar{f}_{ba}) 
\, , \label{ide3} \\
\bar{\BOX} \, \bar{f}_{ab} & = & \frac{2 \Lambda}{(D-1)} \, \bar{f}_{ba} 
\, , \qquad
{\bar{\nabla}}^{a} \, {\bar{\nabla}}_{b} \, \bar{f}_{ac} =
\frac{2 \Lambda}{(D-1)} \, \bar{f}_{bc} \, . \label{ide5}
\end{eqnarray}

Using the Riemann, the Ricci tensors, the Ricci scalar and the Killing-Yano
tensor of rank 2, a covariantly conserved antisymmetric current was
introduced in \cite{kt1}. The `linearized' version of this antisymmetric 
current \footnote{We note in passing that our normalization differs from 
that of \cite{kt1}.}
\be
(j^{ab})_{L} = \bar{f}^{cd} \, (R_{cd}\,^{ab})_{L} 
- 2 \bar{f}^{ac} \, (R_{c}\,^{b})_{L} + 2 \bar{f}^{bc} \, (R_{c}\,^{a})_{L} 
+ \bar{f}^{ab} \, R_{L} \, . \label{jcurl}
\ee
given by \cite{kt1}, is background covariantly conserved. [We emphasize 
that, if the spacetime admits a full Killing-Yano tensor itself, then the 
full $j^{ab}$ is covariantly conserved. However, for the discussion that
follows here, we assume that only the background spacetime has asymptotic 
Killing-Yano tensors.] Then the `linearization' process can be carried out 
in the usual way, by keeping terms linear in the perturbation metric 
$h_{ab}$. Since this current is antisymmetric, this covariant 
conservation leads to an ordinary conservation law via
\be
\bar{\nabla}_{c} ( \sqrt{|\bar{g}|} \, (j^{bc})_{L} ) = 
\partial_{c} ( \sqrt{|\bar{g}|} \, (j^{bc})_{L} ) \, .
\ee
Out of this linearized current, a conserved charge was constructed in
\cite{kt1} for asymptotically transverse \emph{flat} backgrounds, i.e.
they took $\bar{g}_{ab} = \eta_{ab}$. Here we will carry out the analogous
calculation for asymptotically transverse $AdS$ backgrounds.

The crucial step in this computation is to express $(j^{ab})_{L}$ as
the divergence of a completely antisymmetric rank 3 tensor, i.e.
\( (j^{ab})_{L} = {\bar{\nabla}}_{d} \, \bar{\ell}^{abd} \, . \)
Then, up to a trivial normalization, the conserved `charge' can be
obtained as
\be
Q^{ab} \sim \int_{\S} \, dS_{i} \, \sqrt{|\bar{g}|} \, \bar{\ell}^{abi} \, ,
\label{dty}
\ee
where $i$ ranges over the $(D-3)$-dimensional spacelike hypersurface $\S$
at spatial infinity. [One could as well perform similar steps for
asymptotically transverse de Sitter backgrounds, except that one now has to
stay inside the cosmological horizon and assume that the $p$-brane/soliton,
if it at all exists for such backgrounds, does not change the location of 
the horizon.]

In order to find the ``potential'' $\bar{\ell}^{abd}$ for the current, we
need the following expressions for the linearized Riemann, Ricci tensors
and the Ricci scalar:
\begin{eqnarray*} 
(R_{cd}\,^{ab})_{L} & = & \frac{1}{2} 
({\bar{\nabla}}^{a} \, {\bar{\nabla}}_{d} \, h^{b}\,_{c} 
- {\bar{\nabla}}^{b} \, {\bar{\nabla}}_{d} \, h^{a}\,_{c}
- {\bar{\nabla}}^{a} \, {\bar{\nabla}}_{c} \, h^{b}\,_{d}
+ {\bar{\nabla}}^{b} \, {\bar{\nabla}}_{c} \, h^{a}\,_{d} ) \\ 
& & + \frac{\Lambda}{(D-1)(D-2)} ( \d^{a}\,_{d} \, h^{b}\,_{c} 
- \d^{b}\,_{d} \, h^{a}\,_{c} - \d^{a}\,_{c} \, h^{b}\,_{d}
- \d^{b}\,_{c} \, h^{a}\,_{d} ) \, , \\
(R_{c}\,^{b})_{L} & = & - \frac{2 \Lambda}{(D-2)} \, h^{b}\,_{c} + \frac{1}{2}
(- \bar{\BOX} \, h^{b}\,_{c} - {\bar{\nabla}}_{c} \, {\bar{\nabla}}^{b} \, h 
+ {\bar{\nabla}}^{d} \, {\bar{\nabla}}_{c} \, h^{b}\,_{d} 
+ {\bar{\nabla}}^{d} \, {\bar{\nabla}}^{b} \, h_{cd}) \, , \\
R_{L} & = & - \bar{\BOX} \, h 
+ {\bar{\nabla}}^{c} \, {\bar{\nabla}}^{d} \, h_{cd}
- \frac{2 \Lambda}{(D-2)} h \, . 
\end{eqnarray*}
The substitution of these into (\ref{jcurl}) yield
\begin{eqnarray}
(j^{ab})_{L} & = & 
\bar{f}^{cd} \, {\bar{\nabla}}^{a} \, {\bar{\nabla}}_{d} \, h^{b}\,_{c} 
- \bar{f}^{cd} \, {\bar{\nabla}}^{b} \, {\bar{\nabla}}_{d} \, h^{a}\,_{c}
+ \frac{2 \Lambda}{(D-1)(D-2)} (\bar{f}^{ca} \, h^{b}\,_{c} 
- \bar{f}^{cb} \, h^{a}\,_{c}) \nonumber \\
& & + \frac{4 \Lambda}{(D-2)} \bar{f}^{ac} \, h^{b}\,_{c} 
+ \bar{f}^{ac} \, \bar{\BOX} \, h^{b}\,_{c} 
+ \bar{f}^{ac} \, {\bar{\nabla}}_{c} \, {\bar{\nabla}}^{b} h
- \bar{f}^{ac} \, {\bar{\nabla}}^{d} \, {\bar{\nabla}}_{c} \, h^{b}\,_{d}
- \bar{f}^{ac} \, {\bar{\nabla}}^{d} \, {\bar{\nabla}}^{b} \, h_{cd} 
\nonumber \\
& & - \frac{4 \Lambda}{(D-2)} \bar{f}^{bc} \, h^{a}\,_{c}
- \bar{f}^{bc} \, \bar{\BOX} \, h^{a}\,_{c} 
- \bar{f}^{bc} \, {\bar{\nabla}}_{c} \, {\bar{\nabla}}^{a} h
+ \bar{f}^{bc} \, {\bar{\nabla}}^{d} \, {\bar{\nabla}}_{c} \, h^{a}\,_{d}
+ \bar{f}^{bc} \, {\bar{\nabla}}^{d} \, {\bar{\nabla}}^{a} \, h_{cd}
\nonumber \\
& & - \bar{f}^{ab} \, \bar{\BOX} \, h 
+ \bar{f}^{ab} \, {\bar{\nabla}}_{c} \, {\bar{\nabla}}_{d} \, h^{cd}
- \frac{2 \Lambda}{(D-2)} \, h \, \bar{f}^{ab} \, .
\end{eqnarray}
After somewhat lengthy but straightforward calculations, we finally obtain
\be
(j^{ab})_{L} =  3! \, {\bar{\nabla}}_{d} \, 
\Big( \bar{f}^{c[a} \, {\bar{\nabla}}^{b} \, h^{d]}\,_{c} 
+ \frac{1}{2} \, \bar{f}^{[ba} \, {\bar{\nabla}}^{d]} \, h 
+ \frac{1}{2} \, h_{c}\,^{[b} \, {\bar{\nabla}}^{d} 
\, \bar{f}^{|c|a]} 
- \frac{1}{2} \, \bar{f}^{[ba} \, {\bar{\nabla}}_{|c|} 
\, h^{d]c}
+ \frac{1}{6} \, h \, {\bar{\nabla}}^{[b} \, 
\bar{f}^{da]} \Big) \, , \label{yanc}
\ee
where we have used the standard bracket notation to indicate the totally
antisymmetric parts of tensors. We give a rederivation of this result
in Appendix \ref{appa} using the spin-connection and the vielbein 
formulation of gravity.

This current leads to the following conserved `charges'
\begin{eqnarray}
Q^{ab} & = & \frac{3!}{4 \, \Omega_{D-3} \, G_{D}} \, \int_{\S} 
\, dS_{i} \, \sqrt{|\bar{g}|} \, 
\Big( \bar{f}^{c[a} \, {\bar{\nabla}}^{b} \, h^{i]}\,_{c} 
+ \frac{1}{2} \, \bar{f}^{[ba} \, {\bar{\nabla}}^{i]} \, h 
+ \frac{1}{2} \, h_{c}\,^{[b} \, {\bar{\nabla}}^{i} 
\, \bar{f}^{|c|a]} \nonumber \\ 
& & \qquad \qquad \qquad \qquad \qquad \qquad
- \frac{1}{2} \, \bar{f}^{[ba} \, {\bar{\nabla}}_{|c|} 
\, h^{i]c} + \frac{1}{6} \, h \, {\bar{\nabla}}^{[b} \, 
\bar{f}^{ia]} \Big) \, ,
\label{dtyint}
\end{eqnarray}
as an integral over the $(D-3)$-dimensional spatial hypersurface. Our choice
of normalization will be clear from the discussion below. 

This is our main result for conserved rank 2 `Killing-Yano charges' for 
asymptotically transverse $AdS$ spacetimes. Of course, the formula 
(\ref{dtyint}) covers the flat space ($\Lambda = 0$) limit given by 
\cite{kt1}, 
except that our notation is different. Up to now, we have not 
made a specific choice of coordinates, apart from the assumption that the 
perturbation part of the metric behaves in such a way that the `charge' 
integration does not diverge. Thus, as can also be checked explicitly, 
(\ref{dtyint}) is invariant under background diffeomorphisms 
\( \d_\z h_{ab} = {\bar{\nabla}}_{a} \, \z_b + {\bar{\nabla}}_{b} \z_a . \) 

We shall generalize the formula (\ref{dtyint}) to the case of rank $n$ 
Killing-Yano tensors in what follows. However, before we move on to that, 
let us step back and think about the meaning of the current (\ref{jcurl}) 
and the `charge' (\ref{dtyint}) it led to. If anything, it is hard to grasp 
the geometric or the physical meaning of this current. Nevertheless, it 
is certain that being a conserved current, it leads to a conserved `charge', 
whose physical meaning is, unfortunately, again unclear. It was shown 
in \cite{kt1, kt2} that for asymptotically transverse flat spacetimes, 
explicit computation using ``translational Killing-Yano tensors" indicates 
that one gets the ADM mass density. However, this is quite puzzling since 
we know that one gets the ADM mass from Einstein's equation using a 
timelike Killing vector and the symmetric energy momentum tensor $T_{ab}$, 
which is covariantly conserved. Moreover, it is more important to note 
that the meaning of $T_{ab}$ and its relation to geometry is quite clear. 
Keeping this discussion in mind and taking the risk of being pedantic, we 
then ask the following question: Why does one get mass (and angular 
momenta) from the `good old' \emph{symmetric} $T_{ab}$ but, mass (and angular 
momenta) `densities' from completely new \emph{antisymmetric} currents 
like $j_{ab}$? This question, which was not addressed in \cite{kt1, kt2}, 
clearly deserves some attention, whose answer we don't know.

Let us first consider the rather ``unique" rank 1 case and go back to the 
Killing vectors; then the construction above should give the flat space 
ADM \cite{adm} or curved space AD \cite{abdes, des1} charges. Now, instead 
of $\bar{f}^{ba}$, suppose that we have a Killing vector $\bar{\x}^{a}$. 
In this case, (\ref{jcurl}) (or more explicitly (\ref{bigcurrent}) below) 
very naively gives a background conserved current 
\be
(\bar{j}^a)_{L} = - 2 \, \bar{\x}^{b} \, (G^{a}\,_{b})_{L} \, ,
\ee          
which is nothing but (up to a constant and the choice of the sign which is
a matter of conventions) $\bar{\x}^{b} \, (T^{a}\,_{b})_{L}$. The resulting 
charge, whose details were given in \cite{des1} (similar to the form of 
(\ref{dtyint})), reads
\be
Q^{a}(\bar{\x}) = \frac{1}{4 \, \Omega_{D-2} \, G_{D}} 
\int_{\partial \S} \, dS_i \, \sqrt{|\bar{g}|} \, q^{ai} \, , 
\label{admass}
\ee
with
\begin{eqnarray}
q^{ai} & = & \bar{\x}_{b}
{\bar{\nabla}}^{a} h^{ib} - \bar{\x}_{b} {\bar{\nabla}}^{i} h^{ab}
+ \bar{\x}^{a} {\bar{\nabla}}^i h - \bar{\x}^i {\bar{\nabla}}^{a} h
+ h^{ab} {\bar{\nabla}}^i \bar{\x}_{b} \nonumber \\
& & - h^{ib} {\bar{\nabla}}^{a}
\bar{\x}_{b} + \bar{\x}^i {\bar{\nabla}}_{b} h^{ab} - \bar{\x}^{a}
{\bar{\nabla}}_{b} h^{ib} + h {\bar{\nabla}}^{a} \bar{\x}^i
\, , \nonumber \\
& = & 2! \, ( \bar{\x}_{b} \, {\bar{\nabla}}^{[a} \, h^{i]b} +
\bar{\x}^{[a} \, {\bar{\nabla}}^{i]} \, h
+ h^{b[a} \, {\bar{\nabla}}^{i]} \, \bar{\x}_{b}
- \bar{\x}^{[a} \, {\bar{\nabla}}_{|b|} \, h^{i]b}
+ \frac{1}{2} h \, {\bar{\nabla}}^{[a} \, \bar{\x}^{i]} ) \, ,
\label{ad}
\end{eqnarray}
where the integration is over a solid angle of a $(D-2)$-dimensional 
sphere $S^{D-2}$. Comparing (\ref{ad}) with (\ref{dtyint}), we see
how the generic rank $n$ case will read; the underlying procedure is
nothing but a proper antisymmetrization of (\ref{ad}).

The main question now is whether we could have gotten (\ref{dtyint}) 
without referring to the antisymmetric current (\ref{jcurl}) whose 
connection to the physical mass and angular momenta is 
unclear. Of course, the more relevant conserved current to use is 
the energy momentum tensor. Let us show now that one cannot actually 
do that. In fact, consider a \emph{generic} gravity theory 
coupled to a covariantly conserved matter source $\t_{ab}$,
\be
\P_{ab}(g, R, \nabla R, R^2, \dots) = \t_{ab} \, , \label{gengr}
\ee
where $\P_{ab}$ is the ``Einstein tensor'' of a local, invariant
generic gravity action, with \( \P_{ab}(\bar{g}, 
\bar{R}, \bar{\nabla} \bar{R}, \bar{R}^2, \dots) = 0 \)
by assumption. The Bianchi identities of the full theory
(just as the background gauge invariance) are now carried
on to the background such that \( \bar{\nabla}_{a} \P^{ab} = 0. \)

Consider now the following:
\[ \bar{\nabla}_{c} ( \bar{f}^{ab} \, \P_{b}\,^{c} ) =
\bar{\nabla}_{c} ( \bar{f}^{ab} \, \t_{b}\,^{c} ) = 0 \, , \]   
which follows from the background Bianchi identity, (\ref{ide2}) and
the antisymmetry of the Killing-Yano tensor. Note, however, that
\[ \bar{\nabla}_{c} ( \bar{f}^{ab} \, \P_{b}\,^{c} ) =
\partial_{c} ( \bar{f}^{ab} \, \P_{b}\,^{c} ) + \bar{\G}^{a}\,_{cd} \,
\bar{f}^{db} \, \P_{b}\,^{c} + \bar{\G}^{c}\,_{cd} \,   
\bar{f}^{ab} \, \P_{b}\,^{d} \, . \]
Using \( \bar{\G}^{c}\,_{cd} = \partial_{d} \, (\ln{\sqrt{|\bar{g}|}}), \)
this can be written as
\be
\bar{\nabla}_{c} ( \sqrt{|\bar{g}|} \, \bar{f}^{ab} \, \P_{b}\,^{c} ) =
\partial_{c} ( \sqrt{|\bar{g}|} \, \bar{f}^{ab} \, \P_{b}\,^{c} ) +
\sqrt{|\bar{g}|} \, \bar{\G}^{a}\,_{cd} \, \bar{f}^{db} \, \P_{b}\,^{c} \, .
\label{rea}
\ee
The last term on the right hand side of (\ref{rea}) does not necessarily
vanish and this is the reason why an analogous construction as the one 
given in \cite{abdes, des1} with Killing vectors cannot automatically be 
carried on to the case of Killing-Yano tensors, since the tensor density 
current \( \sqrt{|\bar{g}|} \, \bar{f}^{ab} \, \P_{b}\,^{c} \) is not 
\emph{ordinarily} conserved in general. However, if we consider the
antisymmetric combination 
\( \bar{f}^{ab} \, \P_{b}\,^{c} - \bar{f}^{cb} \, \P_{b}\,^{a} \) 
\footnote{We thank D. Kastor for bringing this to our attention.}, 
then we have
\[ \bar{\nabla}_{c} \Big( \sqrt{|\bar{g}|} ( \bar{f}^{ab} \, \P_{b}\,^{c}
- \bar{f}^{cb} \, \P_{b}\,^{a} ) \Big) = 
\partial_{c} \Big( \sqrt{|\bar{g}|} ( \bar{f}^{ab} \, \P_{b}\,^{c}
- \bar{f}^{cb} \, \P_{b}\,^{a} ) \Big) \neq 0 \, , \]
due to the fact that 
\( \bar{\nabla}_{a} ( \bar{f}^{ab} \, \P_{b}\,^{c} ) \neq 0 \). Thus, one is
forced to use the new antisymmetric current (\ref{jcurl}).

For the rank $n$ case, these charges are computed as follows: one starts with 
(\ref{ad}) and first replaces the background Killing vector $\bar{\x}^{a}$ 
with the background Killing-Yano tensor $\bar{f}^{a_1 \dots a_n}$. Next
thing to do is to fully antisymmetrize the expression thus obtained
with respect to the uncontracted indices inside the covariant derivative
in such a way that one gets a fully antisymmetric potential 
$\bar{\ell}^{a_1 \dots a_{n+1}}$. For the rank 2 case, one can easily 
get (\ref{dtyint}) through this method. For the rank $n$ case, we get
\begin{eqnarray}
(\frac{n}{2} \, j^{a_1 \dots a_n})_{L} & = & (n+1)! \, {\bar{\nabla}}_{d} \, 
\Big( \bar{f}^{c[a_1 \dots a_{n-1}} \, {\bar{\nabla}}^{b} \, h^{d]}\,_{c} 
+ \frac{1}{n!} \, \bar{f}^{[b a_1 \dots a_{n-1}} \, {\bar{\nabla}}^{d]} \, h 
+ \frac{1}{n!} \, h_{c}\,^{[b} \, {\bar{\nabla}}^{d} 
\, \bar{f}^{|c| a_1 \dots a_{n-1}]} \nonumber \\
& & \qquad \qquad \qquad - \frac{1}{n!} \, \bar{f}^{[b a_1 \dots a_{n-1}} 
\, {\bar{\nabla}}_{|c|} \, h^{d]c} + \frac{1}{(n+1)!} \, h \, 
{\bar{\nabla}}^{[b} \, \bar{f}^{d a_1 \dots a_{n-1}]} \Big) \, , 
\label{bigch}
\end{eqnarray}
which yields the following conserved `charges' when integrated over a  
$(D-1-n)$-dimensional hypersurface at spatial infinity \footnote{Note 
that the `charges' are gauge/diffeomorphism invariant only at spatial 
infinity.}
\begin{eqnarray}
Q^{a_1 \dots a_n}(\bar{f}) & = & \frac{(n+1)!}{2n \, \Omega_{D-1-n} \, G_{D}} 
\, \int_{\S} \, dS_{i} \, \sqrt{|\bar{g}|} \, 
\Big( \bar{f}^{c[a_1 \dots a_{n-1}} \, {\bar{\nabla}}^{b} \, h^{d]}\,_{c} 
+ \frac{1}{n!} \, \bar{f}^{[b a_1 \dots a_{n-1}} \, {\bar{\nabla}}^{d]} \, h 
\nonumber \\
& & \hspace{-1cm}
+ \frac{1}{n!} \, h_{c}\,^{[b} \, {\bar{\nabla}}^{d} 
\, \bar{f}^{|c| a_1 \dots a_{n-1}]}
- \frac{1}{n!} \, \bar{f}^{[b a_1 \dots a_{n-1}} \, {\bar{\nabla}}_{|c|} 
\, h^{d]c} + \frac{1}{(n+1)!} \, h \, {\bar{\nabla}}^{[b} \, 
\bar{f}^{d a_1 \dots a_{n-1}]} \Big) \, .
\label{esasbigch}
\end{eqnarray}
Once again this formula works both for asymptotically transverse $AdS$ and 
asymptotically transverse flat spacetimes in generic coordinates. We could 
also have gotten this expression using the following conserved current given 
by \cite{kt1} 
\be
j^{a_1 \dots a_n} = (n-1) \, R^{[a_1 a_2}\,_{bc} \, f^{a_3 \dots a_n]bc}
+ 4 (-1)^{n} \, R_{c}\,^{[a_1} \, f^{a_2 \dots a_n]c}
+ \frac{2}{n} \, R \, f^{a_1 \dots a_n} \, .
\label{bigcurrent}
\ee
Obtaining (\ref{esasbigch}) from (\ref{bigcurrent}) by following steps
similar to the ones that took us from (\ref{jcurl}) to (\ref{yanc}) 
(and thus to (\ref{dtyint})) is again straightforward but considerably
longer.

{\bf An explicit example: The long Weyl rod}

As has already been mentioned before, our construction works for
asymptotically transverse flat or $AdS$ spaces. For the latter, we
are not aware of any solution with a \emph{uniform} mass density. However,
for the flat case, we can consider the infinitely extended Weyl rod 
\cite{weyl}. Since this solution is not widely known, let us first write
down the metric for a rod centered at the origin with length $L$ and 
\emph{total} mass $m$:
\be
ds^2 = - e^{2 \p} \, dt^2 + e^{-2 \p} \, [ e^{2 \n} \, (dr^2 + dz^2) 
+ r^2 \, d\vp^2 ] \, ,
\label{rodmet}
\ee
where
\begin{eqnarray}
e^{2 \p} & = & \Big( \frac{R_1+R_2-L}{R_1+R_2+L} \Big)^{2 m/L} \, , \quad
e^{2 \n} = \Big( \frac{\left( R_1+R_2 \right)^2-L^2}{4\;R_1\;R_2} 
\Big)^{4 m^2/L^2} \, , \\
R_1 & = & \sqrt{r^2 + \left( z- \frac{L}{2} \right)^2} \, , \; \quad \quad
R_2 = \sqrt{r^2 + \left( z+ \frac{L}{2} \right)^2} \, .
\end{eqnarray}
The correct background to work with is the flat Minkowski spacetime
written in cylindrical coordinates which is simply obtained by setting
$m=0$. A calculation using (\ref{admass}) with the background Killing
vector $\bar{\x}^{\m} = - \d^{\m}\,_{0}$ indeed yields the mass to be $m$.
[Note that when $L \to -L$, $m \to -m$ and the Schwarzschild solution
can be obtained by setting $m=L$.] As $L \to \infty$, the total mass
obviously diverges. As for the Killing-Yano charges, one starts with a
background Killing-Yano tensor whose only nontrivial component reads
$\bar{f}^{tz}=-1$, and uses (\ref{dtyint}). The result is $Q^{tz} = m/L$, 
which is simply the constant mass density.

\section{Conclusions and discussions}

Using the antisymmetric background Killing-Yano tensors of rank $n$, we
have constructed conserved gravitational charges of $D$-dimensional 
spacetimes that asymptotically approach to Anti-de Sitter or flat spaces 
only in some transverse directions of an extended gravitational solution. 
This work generalizes that of Kastor-Traschen \cite{kt1} which dealt only 
with asymptotically transverse flat spacetimes. Our construction 
is based on the linearization of a completely antisymmetric current 
(\ref{bigcurrent}) given by \cite{kt1}, whose physical meaning is 
somewhat unclear. This current makes no reference to a specific gravity 
model. 

The conserved charge densities are background diffeomorphism invariant. It
is somewhat surprising that one can get anything meaningful out of 
Killing-Yano tensors with regard to the conserved quantities of ``extended" 
gravitational solutions, such as $p$-branes. To begin with, a naive 
application of the ADM or AD charges gives divergent results for the total 
mass and the total angular momenta. However, the mass density and the angular 
momentum density turn out to be finite which can be expected. What 
is quite interesting is that these ``densities" are expressed as 
integrals over a $(D-p-2)$-dimensional hypersurface at spatial infinity, 
in close analogy with the ADM or AD case. In fact, the main difference is 
that, for Killing-Yano charges parallel directions to the ``brane" are 
left alone in the integration process.

As already mentioned, these conserved gravitational charges, constructed out 
of the Killing-Yano tensors, turn out to be ``charge (mass) densities". One
might simply ask why one would bother with these conserved quantities at 
the first place when one can simply go back to the ADM or the AD formulas 
and define charge densities out of these. This is, of course, a 
legitimate question which has two possible answers: Firstly, a naive mass 
density definition obtained from the ADM or the AD expression would not 
be diffeomorphism invariant, the use of Killing-Yano tensors renders the 
result coordinate invariant. This is important. Secondly, it is better to 
read our result from a different perspective. We have established the 
following: If a spacetime asymptotically admits antisymmetric 
Killing-Yano tensors, the conserved quantities they generate are the mass 
and the angular momentum densities.

As for an explicit example, we have given the mass density of the axisymmetric
Weyl rod in $D=3+1$ dimensions. It would, of course, be better to give 
more examples, especially consider those solutions which are transverse
asymptotically $AdS$ and those that also involve rotation, but unfortunately 
we are not aware of such solutions.
                                                                              
Finally, we have especially refrained from a discussion on the positivity
of these ``mass-densities" (for the asymptotically transverse flat spacetimes 
case, such a discussion was given in \cite{kt2}). In asymptotically $AdS$ 
spacetimes, there are negative mass objects (see e.g. \cite{ceb} and the 
references therein). Thus, in general, we do not expect `a positive mass
theorem' to hold in these spacetimes. It is true that, following e.g.
Gibbons et al. \cite{gib}, one can presumably show the positivity of mass 
density in certain 4-dimensional spacetimes by rewriting the mass density 
in terms of the square of certain Witten-Nester spinors. However, for many 
interesting solutions, such as the $AdS$ soliton, one does not have regular 
spinors \cite{sken} satisfying the requirements of positivity.

\appendix
\section{\label{appa} Derivation of the Killing-Yano charges in the 
spin-connection formalism}

In this appendix, we will redrive (\ref{yanc}) using the first order form
of gravity. This is not a redundant task at all, since it maybe relevant 
to various supergravity models.

Analogous to the definition of a Killing-Yano tensor of rank 2, we define 
a Killing-Yano 2-form $f$ in terms of orthonormal co-frame 1-forms $e^{a}$ as 
\be
f = \frac{1}{2} \, f_{ab} \, e^{a} \we e^{b} \, . \label{1}
\ee
The current (\ref{bigcurrent}) for the case $n=2$ now can be written as
\be
j = f_{ab} \, R^{ab} + 2 \io_{a} f \we P^{a} + f R \, , \label{2}
\ee
where $R^{ab}$ is the curvature 2-form defined in the usual way as
\be
R^{ab} = d \omega^{ab} + \omega^{a}\,_{c} \we \omega^{cb}
\label{3}
\ee
in terms of the connection 1-forms $\omega^{a}\,_{b}$; $P^{a}$ stands for
the Ricci 1-forms defined through $P^{a} = \io_{b} R^{ba}$ and $R$ is
the curvature scalar $R = \io_{a} \io_{b} R^{ba}$. The current $j^{ab}$
can be obtained in an orthonormal frame as
\be
j^{ab} = f_{cd} \, R^{abcd} - 2 f^{a}\,_{c} \,
P^{cb} + 2 f^{b}\,_{c} \, P^{ca} + f^{ab} \, R \, , \label{6}
\ee
by noting that
\be
P^{a} = P^{a}\,_{b} e^{b} \,, \qquad 
R^{ab} = \frac{1}{2} R^{ab}\,_{cd} \, e^{c} \we e^{d} \quad \mbox{and} \quad
j = \frac{1}{2} j_{ab} \, e^{a} \we e^{b} \, ,
\label{4}
\ee
where $f_{ba} = \io_{a} \io_{b} f$. 

Suppose also that the metric tensor $g = \eta_{ab} \, e^{a} \otimes e^{b} $ 
is decomposed such that the `full' orthonormal coframe 1-form $e^{a}$ 
can be written as the sum of a `background' orthonormal coframe 1-form
$\bar{e}^{a}$ plus a `deviation' piece as
\be
e^{a} \equiv \bar{e}^{a} + \vp^{a}\,_{b} \, \bar{e}^{b} \, ,
\label{7}
\ee
where the 0-forms $\vp^{a}\,_{b}$ are as usual assumed to vanish
sufficiently rapidly at the `spatial infinity'. The background spacetime 
geometry is assumed to satisfy the cosmological Einstein field equation
\be
\bar{R}^{ab} \we \bar{\ast} ( \bar{e}_{a} \we \bar{e}_{b}
\we \bar{e}_{c} ) = \Lambda \, \bar{\ast} \, \bar{e}_{c} \, ,
\label{8}
\ee
where $\bar{\ast}$ is the background Hodge operator. (\ref{8}) are solved 
by a space of constant curvature which satisfies
\begin{eqnarray}
\bar{R}_{abcd} & = & \frac{2 \Lambda}{(D-1)(D-2)} \, ( \eta_{ac} \, \eta_{bd} -
\eta_{ad} \, \eta_{bc} ) \, , \nonumber \\
\bar{R}_{ab} & = & \frac{1}{2} \, \bar{R}_{abcd} \, \bar{e}^{c} 
\we \bar{e}^{d} = \frac{2 \Lambda}{(D-1)(D-2)} \, \bar{e}_{a} \we \bar{e}_{b} 
\, , \label{9} \\
\bar{P}_{a} & = & \frac{2 \Lambda}{D-2} \bar{e}_{a} \, , \quad
\bar{P}_{ab} = \frac{2 \Lambda}{D-2} \eta_{ba} \, , \quad
\bar{R} = \bar{\io}_{a} \, \bar{\io}_{b} \, \bar{R}^{ba} =
\frac{2 \Lambda D}{D-2} \, . \nonumber
\end{eqnarray}
Moreover, the relation $\io_{b}\, e^{a} = \d_{b}\,^{a}$ implies that
\( \io_{b} = \bar{\io}_{b} - \vp_{bs} \bar{\io}^{s} \)
in terms of the inner product operator $\bar{\io}_{b}$ of the
background spacetime that satisfies 
\( \bar{\io}_{b} \, \bar{e}^{a} = \d_{b}\,^{a} \). 
Now the Cartan structure equations [in the case of vanishing torsion $T^{a}$]
\( d e^{a} + \omega^{a}\,_{b} \we e^{b} = 0 \) yield 
\be
\omega^{a}\,_{b} = \bar{\omega}^{a}\,_{b} + \bar{e}^{c} \left( \bar{\io}_{b} D 
\vp^{a}\,_{c} - \bar{\io}^{a} D \vp_{cb} \right) \label{12}
\ee
to first order (linearized form) in $\vp^{a}\,_{b}$, where for any 
$p$-form $\Omega^{a}\,_{b}$
\be
D \Omega^{a}\,_{b} = d \vp^{a}\,_{b} + \bar{\omega}^{a}\,_{c}
\we \Omega^{c}\,_{b} - \bar{\omega}^{c}\,_{b} \we \Omega^{a}\,_{c} \label{13}
\ee
and $\bar{\omega}^{a}\,_{b}$ satisfies the background Cartan structure
equation
\( d \bar{e}^{a} + \bar{\omega}^{a}\,_{b} \we \bar{e}^{b} = 0 . \)
Using (\ref{3}) and these preliminaries, the linearized curvature 
2-form can be calculated as
\be
R^{ab} = \bar{R}^{ab} - \bar{e}^{c} \we D (\bar{\io}^{b} D
\vp^{a}\,_{c} ) + \bar{e}^{c} \we D( \bar{\io}^{a} D \vp_{c}\,^{b} ) \, ,
\label{15}
\ee
which yields the following formulas for the Ricci 1-form
\be
P^{a} = \bar{P}^{a} - D ( \bar{\io}^{a} D \vp^{b}\,_{b} )
+ D ( \bar{\io}^{b} D \vp_{b}\,^{a} ) + \bar{\io}_{b} D
(\bar{\io}^{a} D \vp^{b}\,_{c} ) \bar{e}^{c} 
- \bar{\io}_{b} D ( \bar{\io}^{b} D \vp_{c}\,^{a} )
\bar{e}^{c} - \vp_{bs} \bar{\io}^{s} \bar{R}^{ba} \, ,
\label{16}
\ee
and the curvature scalar
\be
R = \bar{R} - 2 \bar{D}_{b}\, \bar{D}^{b} \vp_{a}\,^{a} + 2
\bar{D}_{a}\, \bar{D}_{b} \vp^{ab} - 2 \vp_{ab}\, \bar{P}^{ab} 
\, , \label{17}
\ee
where \( \bar{D}_{b} \vp^{ba} = \bar{\io}_{b} D \vp^{ba} \)
and \( \bar{D}_{a}\, \bar{D}_{b} \vp^{ba} = \bar{\io}_{a} D (
\bar{\io}_{b} D \vp^{ba} ) \).

The use of the background Killing-Yano 2-form
\( f = \frac{1}{2} \bar{f}_{ab} \bar{e}^{a} \we \bar{e}^{b} \), 
and the substitution of (\ref{15}), (\ref{16}) and (\ref{17}) into 
(\ref{2}) gives the following expression for the linearized current 
2-form $j_{L}$:
\begin{eqnarray}
j_{L} & = & \bar{f}_{ab} \bar{e}^{c} \we D 
( \bar{\io}^{a} D \vp_{c}\,^{b} ) - \bar{f}_{ab} \bar{e}^{c}
\we D ( \bar{\io}^{b} D \vp^{a}\,_{c} ) - \vp_{an}
\bar{f}^{n}\,_{b} \bar{R}^{ab} - \vp_{bl} \bar{f}_{a}\,^{l}
\bar{R}^{ab} \nonumber \\
& & - 2 \bar{\io}_{a} f \we D ( \bar{\io}^{a} D
\vp^{b}\,_{b} ) + 2 \bar{\io}_{a} f \we D ( \bar{\io}^{b} D 
\vp_{b}\,^{a} ) + 2 \bar{\io}_{a} f \we \bar{e}^{c} \bar{\io}_{b} D 
( \bar{\io}^{a} D \vp^{b}\,_{c} ) \nonumber \\
& & - 2 \bar{\io}_{a} f \we \bar{e}^{c}
\bar{\io}_{b} D ( \bar{\io}^{b} D \vp_{c}\,^{a} ) - 2
\vp_{bs} \bar{\io}_{a} f \we \bar{\io}^{s} \bar{R}^{ba}
- 2 \vp_{as} \bar{\io}^{s} f \we \bar{P}^{a} \nonumber \\
& & + 2 f \bar{\io}_{a} D\, ( \bar{\io}_{b} D
\vp^{ab} ) - 2 f \bar{\io}^{b} D\, ( \bar{\io}_{b} D
\vp_{a}\,^{a} ) - 2 f \vp_{ab} \bar{P}^{ab} \, .
\label{22}
\end{eqnarray}
This linearized current can be written in terms of the
background orthonormal co-frames as 
\( j_{L} = \frac{1}{2} (j^{cs})_{L} \, \bar{e}_{c} \we \bar{e}_{s} \):
\begin{eqnarray}
(j^{cs})_{L} & = & \bar{f}_{ab} \left(
\bar{D}^{s} \bar{D}^{a} \vp^{cb} - \bar{D}^{c} \bar{D}^{a}
\vp^{sb} + \bar{D}^{c} \bar{D}^{b} \vp^{sa} - \bar{D}^{s}
\bar{D}^{b} \vp^{ca} \right) \nonumber \\
& & - \vp_{al} \bar{f}^{l}\,_{b} \bar{R}^{abcs} -
\vp_{bl} \bar{f}_{a}\,^{l} \bar{R}^{abcs} 
+ 2 \left( \bar{f}_{a}\,^{c}
\vp_{bl} \bar{R}^{abls} - \bar{f}_{a}\,^{s} \vp_{bl}
\bar{R}^{ablc} \right) \nonumber \\
& & + 2 \bar{f}_{a}\,^{c} \left( \bar{D}^{s} \bar{D}^{b} \vp_{b}\,^{a}
- \bar{D}^{s} \bar{D}^{a} \vp^{b}\,_{b} 
+ \bar{D}_{b} \bar{D}^{a} \vp^{sb} 
- \bar{D}^{b} \bar{D}_{b} \vp^{sa} \right) \nonumber \\
& & - 2 \bar{f}_{a}\,^{s} \left( \bar{D}^{c} \bar{D}^{b} \vp_{b}\,^{a} 
- \bar{D}^{c} \bar{D}^{a} \vp^{b}\,_{b}
+ \bar{D}_{b} \bar{D}^{a} \vp^{bc} - \bar{D}^{b} 
\bar{D}_{b} \vp^{ca} \right) \nonumber \\
& & - 2 \left( \vp_{al} \bar{f}^{lc} \bar{P}^{as} -
\vp_{al} \bar{f}^{ls} \bar{P}^{ac} \right) + 2 \bar{f}^{cs}
\left( \bar{D}_{a} \bar{D}_{b} \vp^{ba} - \bar{D}^{b}
\bar{D}_{b} \vp_{a}\,^{a} - \vp_{ab} \bar{P}^{ab} \right) .
\label{23}
\end{eqnarray}
The next thing to do is to express (\ref{23}) in the form
\( (j^{cs})_{L} = \bar{D}_{a} ( l^{acs} ) \), where $l^{acs}$ is totally
antisymmetric in its $c$, $s$ and $a$ indices.

Finally, using the following identities
\begin{eqnarray}
\bar{D}_{c} \bar{D}_{d} \bar{f}_{ba} & = & - \frac{1}{2} \left(
\bar{R}^{n}\,_{cdb} \bar{f}_{an} + \bar{R}^{n}\,_{cad}
\bar{f}_{bn} + \bar{R}^{n}\,_{cba} \bar{f}_{dn} \right) \, , \label{25} \\
( \bar{D}_{c} \bar{D}_{d} - \bar{D}_{d} \bar{D}_{c} ) \vp_{ba}
& = & - \bar{R}^{l}\,_{bcd} \vp_{la} - \bar{R}^{l}\,_{acd} \vp_{bl} 
\, , \quad
\bar{D}^{b} \bar{D}_{b} \bar{f}_{ca} = - \frac{2 \Lambda}{D-1}
\bar{f}_{ca} \, , \label{26}
\end{eqnarray}
(\ref{23}) can be written as 
\be
(j^{cs})_{L} = 2 \, 3! \, \bar{D}_{a} \Big( \bar{f}^{b[a}
\bar{D}^{c} \vp^{s]}\,_{b} + \frac{1}{2} \bar{f}^{[ca} \bar{D}^{s]}
\vp^{b}\,_{b} 
+ \frac{1}{2} \bar{D}_{b} \bar{f}^{[ca} \vp^{s]}\,_{b} 
- \frac{1}{2} \bar{f}^{[ca} \bar{D}_{b} \vp^{s]b}
+ \frac{1}{6} \vp^{b}\,_{b} \bar{D}^{[a} \bar{f}^{cs]} \Big) 
\, , \label{30}
\ee
which corresponds to (\ref{yanc}).

\begin{acknowledgments}

We would like to thank J. Jezierski for a useful discussion on Killing-Yano
tensors of $AdS$ spacetimes. We would also like to thank D. Kastor for
pointing out a serious mistake in the earlier version of this manuscript. 
This work is partially supported by the Scientific and Technological Research 
Council of Turkey (T{\"U}B\.{I}TAK). B.T. is also partially supported by 
the ``Young Investigator Fellowship" of the Turkish Academy of Sciences 
(T{\"U}BA) and by the  T{\"U}B\.{I}TAK Kariyer Grant 104T177. 

\end{acknowledgments}

\end{document}